\documentclass[cameraready]{Interspeech}

\title{What Do Deepfake Speech Detectors Actually Hear?}

\author[orcid=0009-0000-5722-0571]{Vojtěch}{Staněk}
\author{Veronika}{Jirmusová}
\author[orcid=0000-0002-4717-1910]{Anton}{Firc}
\author[orcid=0000-0002-9009-2193]{\\Kamil}{Malinka}
\author[orcid=0009-0004-6055-5136]{Jakub}{Reš}
\author[orcid=0000-0002-2875-9567]{Martin}{Perešíni}

\address{
    Security@FIT, Brno University of Technology, Czech Republic
}

\email{istanek@fit.vut.cz, xjirmu00@vut.cz, ifirc@fit.vut.cz,\\malinka@fit.vut.cz, iresj@fit.vut.cz, iperesini@fit.vut.cz}


\usepackage{comment}
\usepackage{amsmath}   
\usepackage{amsfonts}  
\usepackage{amssymb}  
\usepackage{hyperref}
\usepackage{multirow}
\usepackage{adjustbox}
\usepackage[table]{xcolor}

\begin{document}

\maketitle

\begin{abstract}
    Deepfake speech detectors often output a single score without explaining \textit{why} an audio sample is flagged, \textit{where} in the signal the evidence lies, or \textit{what} cues drive the decision. We propose an audio-native explainability pipeline using Integrated Gradients on time-aligned self-supervised representations to localize decision evidence over time. We apply the proposed method to three WavLM-based detectors (AASIST, CA-MHFA, SLS) on ASVspoof~5 and manually annotate the highest-attribution regions to provide a semantic meaning of the most important cues. Despite similar performance, the detectors rely on different cues: AASIST emphasizes non-speech/environment cues, CA-MHFA focuses on localized phoneme artifacts, and SLS relies on word boundaries and spectral integrity. We move beyond speculative reasoning and validate our findings by causal masking of the primary detector cues. Observed performance degradation further supports the explained detector semantics.
\end{abstract}

\section{Introduction}


High-quality speech deepfakes are now easy to generate. In response, deepfake speech detectors are employed as the primary defense~\cite{wang24_asvspoof, FircSpectrogram, MalinkaVoiceAssistants}. 
A key limitation is that these systems typically output a single score~\cite{aliyev24_asvspoof, chan24_asvspoof}, with little insight into \textit{why} a recording received the score, \textit{where} in the utterance the detector finds evidence, or \textit{what} type of cue it relies on, especially for self-supervised (SSL) pipelines that dominate recent 
systems~\cite{sls, stourbe24_asvspoof}.

In this work, we study the explainability (XAI) of what modern deepfake speech detectors rely on by analyzing three current WavLM-based architectures: AASIST~\cite{jung22aasist, borodin24_asvspoof}, Context-Aware MHFA (CA-MHFA)~\cite{BUT198050, rohdin24_asvspoof}, and Sensitive Layer Selection (SLS)~\cite{sls}. The examined detectors represent modern systems (EER 3.98\%-5.26\% on ASVspoof~5~\cite{wang24_asvspoof}; see \autoref{tab:fusion_results}) with competitive performance compared to the current single-system state-of-the-art (EER 3.75\%)~\cite{peng2025hybridpruning}. 

We explain the detectors' decisions using an audio-native adaptation of Integrated Gradients (IG)~\cite{sun} on time-aligned SSL frame representations. To move beyond isolated qualitative examples, we conduct a structured annotation and evaluation of regions with the highest IG attributions and provide a characterization of consistent cues and failure patterns across detectors. While prior analyses typically remain qualitative, our approach assigns semantic meaning to \textit{what} drives each of the examined detectors, supported by a causal masking validation.

Our results on ASVspoof 5 show that detectors with similar performance can learn significantly different decision logic: AASIST often emphasizes non-speech and environmental regions, CA-MHFA focuses on highly localized phoneme artifacts, and SLS concentrates on word boundaries and spectral integrity cues. These findings provide a practical answer to what deepfake speech detectors actually learn and where they look when producing a decision score. Ultimately, this interpretability empowers forensic analysts to justify detection scores, supports informed detector design choices, and helps anticipate and mitigate failures in deployment.
Our contributions are:

\begin{itemize}[itemsep=-0.5pt, topsep=0pt]
    \item We introduce an \textbf{audio-native adaptation} of IG to explain SSL-based detectors and localize evidence in time.
    \item We perform \textbf{semantic analysis} of AASIST, CA-MHFA, and SLS on ASVspoof~5 to characterize each detector's focus and assign semantic meaning to the identified patterns.
    \item We \textbf{causally validate} the identified primary cues of the detectors via targeted cue masking.
\end{itemize}

\section{Background}

Current deepfake detection systems rely on architectures that use SSL front-ends~\cite{Firc2025}. Pretrained models like Wav2Vec 2.0~\cite{Baevski_wav2vec2, xu24_asvspoof} and WavLM~\cite{stourbe24_asvspoof, wavlm} are favored for their ability to extract rich speaker representations directly from raw audio. These features are processed most often by AASIST~\cite{jung22aasist, xia24_asvspoof, diffuse}, SLS~\cite{sls, scdf}, or (CA-)MHFA~\cite{BUT198050, rohdin24_asvspoof, evolutionary_fusion}. Architectures also often incorporate ResNet-based features~\cite{dao24_asvspoof} or ECAPA-TDNN~\cite{kulkarni24_asvspoof} blocks to enhance channel attention. While these deep networks generalize well by capturing complex dependencies, their structure creates black-box opacity, requiring rigorous interpretability methods to validate decision logic~\cite{Firc3}.

To address this opacity, Explainable AI (XAI) has emerged as a critical component in audio forensics. In speech processing, post-hoc methods like SHAP~\cite{SHAP, ge2024explainingdeeplearningmodels} and LIME~\cite{ribeiro2016why} are frequently employed. Within deepfake speech detection, prior work relied on Class Activation Mapping (CAM)~\cite{ge22_odyssey} or attention visualization~\cite{sls, xia24_asvspoof}. In contrast, Integrated Gradients~\cite{sun} offer an axiomatic alternative, avoiding hard masking by integrating from a bona fide centroid baseline to the input. 

\section{Method}

To interpret the decision-making logic of deepfake detectors, we employ the Integrated Gradients (IG)~\cite{sun} method. 
IG enables precise attribution of decision scores, which is crucial for distinguishing whether a model detects genuine synthesis artifacts or merely exploits channel-specific noise.

\subsection{Integrated Gradients for SSL Representations}

Given an input recording $r$, the front-end SSL encoder (WavLM base+)~\cite{wavlm} produces hidden representations $\mathbf{H} \in \mathbb{R}^{L \times T \times D}$, where $L$ represents the number of transformer layers (including the output of the convolutional encoder), $T$ is the number of time frames, and $D$ is the feature dimension. The back-end detector can be formally defined as $F: \mathbb{R}^{L \times T \times D} \to \mathbb{R}$, which maps these representations to a scalar spoofing score (logit $z$).

IG assigns an attribution to each element $h_{i} \in \mathbf{H}$ by integrating the gradient of $z$ along the straight-line path from a \textit{baseline} $\mathbf{H}'$ (described in \autoref{subsec:baseline}) to the input $\mathbf{H}$:
\begin{equation*}
\text{IG}_{i}(\mathbf{H}, \mathbf{H}') = (h_{i} - h'_{i}) \int_{\alpha = 0}^{1} \frac{\partial F(\mathbf{H}' + \alpha(\mathbf{H} - \mathbf{H}'))}{\partial h_{i}} \, d\alpha
\end{equation*}
where $i$ denotes the feature dimension index and $\alpha$ is the interpolation step. We leverage the \texttt{Captum} library, which approximates this integral with a Riemann sum. 
We compute the attributions with respect to $\mathbf{H}$ targeting the logit $z = F(\mathbf{H})$. 
Since the WavLM convolutional encoder operates with a stride of 320 samples, the time dimension $T$ corresponds to overlaying frames of approximately 20\,ms (at 16kHz). 

\subsection{Temporal Attribution Map}

IG yields attributions $\mathrm{IG}_{l,t,d}$ for each SSL embedding element. Since our goal is to localize \textit{where} in the utterance the detector's focus lies, we sum the attribution scores across all $L$ layers and $D$ features to obtain a unified temporal map $A_t \in \mathbb{R}^T$:
\begin{equation*}
A_t = \sum_{l=1}^{L} \sum_{d=1}^{D}  \text{IG}_{l,t,d}(\mathbf{H}, \mathbf{H}').
\end{equation*}
This produces a time-attribution signal in which large positive values correspond to evidence that increases the spoof score, while large negative values correspond to evidence that increases the bona fide score. To mitigate frame-level noise, we smooth $A_t$ using a sliding window average of 6 frames ($\approx 120$\,ms). Both the raw and smoothed attributions are used to identify primary cue regions in our annotation protocol.

\subsection{Annotation Protocol}
\label{subsec:annotation}

To evaluate the semantic relevance of the generated attributions, we establish a structured annotation protocol using a custom web-based interface. Human annotators analyze both the smoothed and raw attribution maps $A_t$ alongside the audio signal and its STFT spectrogram. The protocol has three stages:
\begin{enumerate}[itemsep=0pt, topsep=0pt]
    \item The analyzed recording and its spectrogram are presented to the annotators for inspection. The annotators can input a general observation about the recording.
    \item A series of annotations is gathered for each of the detectors.
    \begin{itemize}[itemsep=0pt, topsep=0pt]
        \item \textbf{Primary cue:} Annotators identify a primary segment that corresponds to the highest IG attributions.
        \item \textbf{Cue type:} Each identified cue is assigned a type from a predefined set: local glitch, phoneme content/articulation, voiced-unvoiced transition, silence, breath, channel/codec noise, spectral artifact, unclear/diffuse attribution.
        \item \textbf{Locality:} Assessing how concentrated or spread is the attribution (Likert scale 1-5~\cite{itu-t}).
        \item \textbf{Qualitative Comment:} A brief description of the specific nature of the cue (optional).
    \end{itemize}
    \item Finally, for the same recording, similarity or disparity between the cues identified by different detectors is assessed.
\end{enumerate}

The protocol is designed to both localize the dominant evidence driving the detector's decision and assign an interpretable semantic label to that evidence. 
Our method is not model-agnostic and focuses on SSL-based models. But, importantly, it enables analysis of the detector semantics and can be generalized to other models operating on time-frame representations.

\section{Experimental Setup}

\begin{figure*}[htbp]
    \centering
    \includegraphics[width=0.94\linewidth]{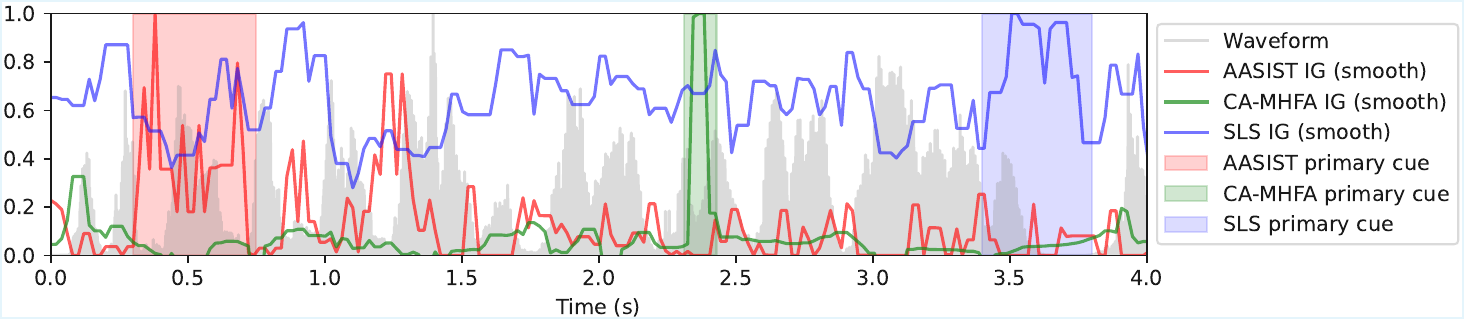}
    \caption{Spoof IG attributions (smoothed) of the three examined detectors for the file E\_0005076209 (spoofed recording from the High-confidence correct predictions category). The highlighted regions represent the primary cue regions identified by one of the annotators.}
    \label{fig:ig}
    \vspace{-1.6em}
\end{figure*}


We evaluate three modern deepfake detection architectures: AASIST~\cite{jung22aasist}, Context-Aware MHFA (CA-MHFA)~\cite{BUT198050}, and Sensitive Layer Selection (SLS)~\cite{sls}. All systems utilize the pre-trained WavLM Base+ model as the front-end feature extractor. To ensure optimal adaptation, we employ a joint training strategy in which both the SSL front-end and the classifier are fine-tuned simultaneously for 10 epochs. We use the AdamW optimizer with a learning rate of $10^{-6}$ for the WavLM parameters and $10^{-3}$ for the classifier parameters with a batch size of 16. We apply stochastic data augmentation with 30\% chances of applying each of the following: time masking, mu-law encoding/decoding, RawBoost (LnL-ISD)~\cite{RawBoost}, adding noise, and one of several filter types (band-pass, high/low-pass, shelf, peaking). Experiments are conducted on the ASVspoof 5 dataset and use the Equal Error Rate (EER) and minDCF performance metrics~\cite{wang24_asvspoof}. 
Code and implementation details are available in a GitHub repository\footnote{\tiny \url{https://github.com/Security-FIT/IG_for_SSL_detectors}}.


\subsection{Baseline Implementation}
\label{subsec:baseline}

The \textit{baseline} $\mathbf{H}'$ is computed individually for each system. Since the SSL front-end is fine-tuned jointly, the resulting feature spaces differ across architectures. We derive a model-specific centroid $\boldsymbol{\mu}_{l,d} \in \mathbb{R}^{L \times D}$ by averaging bona fide features across all $N$ training samples and $T$ time steps:
\begin{equation*}
\boldsymbol{\mu}_{l,d} = \frac{1}{N \cdot T} \sum_{n=1}^{N} \sum_{t=1}^{T} \mathbf{H}_{n,l,t,d}
\end{equation*}
This centroid $\boldsymbol{\mu}_{l,d}$ is then broadcast across the temporal dimension to form the baseline $\mathbf{H}'$, such that for every time step $t \in \{1, \dots, T\}$, the representation remains constant.

Choosing the bona fide centroid as a reference is crucial for SSL-based encoders. Since these models are pre-trained on vast amounts of speech, an auxiliary baseline, such as a zero vector $\mathbf{0}$, can represent an out-of-distribution input that introduces undesired artifacts. We also considered alternative baselines, such as a centroid computed over the entire data set (including spoofed samples) or a noise-only vector. However, a global centroid would introduce bias from specific spoofing attacks present in the training partition, potentially masking the very artifacts we aim to detect. Similarly, a noise-only baseline fails to provide a meaningful semantic reference for speech analysis. By referencing against the bona fide centroid, we ensure that the resulting attributions strictly isolate the deviation caused by spoofing artifacts and digital processing, effectively highlighting what distinguishes a deepfake from natural speech.


\subsection{Subset Selection for Interpretability}

To conduct a rigorous, granular analysis, we curated a representative subset of 100 recordings from the ASVspoof 5 evaluation partition. The selection process was divided into three categories based on model consensus and prediction confidence:
\begin{enumerate}
    \item \textit{High-confidence correct predictions}: First, we identified samples where all three detectors correctly classified the input with high certainty (filtering for correct decisions at an operating point of FAR/FRR $<0.1$\% for all three detectors). From this pool, we selected 32 spoofed recordings (2 samples per attack algorithm) and 28 bona fide recordings.

    \item \textit{High-confidence errors}: To investigate severe model failures, we selected 20 samples where all three detectors were confidently wrong (similarly as above). 
    This subset includes 10 bona fide samples misclassified as spoof (false reject) and 10 spoofed samples misclassified as bona fide (false accept).

    \item \textit{Boundary cases}: 
    We compute the EER threshold for each detector on the dev set, then select evaluation recordings near these thresholds for which all three detectors agree on the predicted label. From this pool, we sample 5 recordings from each outcome category (TP, TN, FP, FN).
    
\end{enumerate}

This selection strategy yields a subset of 100 recordings, balancing clear cases with borderline decisions, providing comprehensive coverage for the subsequent IG analysis. Annotations were completed by three researchers with deepfake speech expertise in a quiet setting using headphones, requiring approximately 40 person-hours in total.
The complete list of selected recordings and the annotation methodology details are available in the GitHub repository.

\section{Results}
\label{sec:results}

Across the 100 examined recordings, we observe two dataset-level patterns: (1) most high-confidence errors are associated with aggressive compression; (2) the attack A28 (pretrained YourTTS~\cite{casanova2022yourtts}) dominates the high-confidence error category. Regarding the examined models, the IG annotations show that the three detectors rely on distinct cues, effectively acting as complementary observers, as visible in \autoref{fig:ig}. The qualitative findings presented below are grounded in the structured annotations of the primary cue regions (\autoref{subsec:annotation}). Specifically, we characterize detector behavior by aggregating the identified cue types, locality scores, and annotator comments across the annotated subset.


\subsection{Primary Focus of Detectors}


\textbf{AASIST} primarily evaluates non-speech and environmental cues. Rather than analyzing what is present in the speech signal, it frequently evaluates what is missing. It assigns high attribution to low-energy regions, often flagging unusually clean silence as a strong deepfake indicator. Otherwise, AASIST's attributions are usually scattered across the recording (Locality 1 or 2). Attribution peaks are usually tied to abrupt changes in the noise profile (e.g., sharp cuts to silence or abrupt sentence starts), indicating sensitivity to unnatural editing and channel artifacts. Because AASIST monitors the overall acoustic environment, it is highly susceptible to misinterpreting low-quality audio, frequently attributing widespread codec artifacts in bona fide recordings as spoof.


In contrast, \textbf{CA-MHFA} acts as a highly localized detector focused on articulation and phonemes. Attributions are predominantly spiky (Locality 4-5), highlighting short segments within speech such as robotic phonemes, consonant bursts, and unnatural phonetic transitions. It is particularly sensitive to sibilants and fricatives (e.g., ``s" phonemes). Furthermore, CA-MHFA focuses on speech fluency, marking abrupt word onsets/offsets as spoof cues. Conversely, it relies on natural connections, reverb, and word continuation, with accented ending consonants being significant contributions to bona fide attributions.


\textbf{SLS} emphasizes global signal integrity and spectral consistency. Its attributions are often spread (Locality 1), with peaks at transition events such as unnatural cuts, artificial cutoffs, and word endings that lack natural reverb. It also responds to broad spectral anomalies (e.g., missing frequency bands or unnatural high-frequency energy). SLS also shows higher attribution in voiced segments than in silent or unvoiced regions.

\textbf{Disparity and Complementarity}: The detector-similarity evaluation shows a lack of consensus among detectors. The models often correctly flag a spoof, but for entirely different reasons. For example, in recording E\_0005076209 (\autoref{fig:ig}), AASIST flagged a ``s" fricative, CA-MHFA pinpointed a specific ``e" phoneme, and SLS targeted word offset and silence. Even when evaluating a similar region, the models appear to capture different phenomena. Agreement across models occurs mainly in two cases. Firstly, heavy compression removes the disparity, as all three models evaluate background noise and misclassify the file based on extreme compression artifacts. Secondly, the models occasionally align on strong bona fide cues, such as a natural, breathy speaking style or fluent word transitions and prominent ending consonants. Overall, the low overlap of primary cues suggests that the detectors rely on distinct representations of synthetic speech and are therefore complementary (see \autoref{fig:ig}). This provides strong empirical evidence that ensemble methods could benefit from combining these models, as the learned factors do not naturally overlap unless forced by heavy compression. To validate, we train a simple logistic regression fusion on the three detector scores on the development subset and observe improved evaluation performance in \autoref{tab:fusion_results}. The performance gain is modest, likely because the detectors share failure modes (primarily due to compression), further suggesting that targeted mitigation of common weaknesses could generally improve all three systems.

\textbf{Natural Breathing as a Bona Fide Marker}: The manual annotations uncovered that respiratory sounds are a strong marker for the bona fide class and one of the few phenomena where detectors often agree. When a natural breath is present, it frequently generates strong bona fide attributions, whereas distorted, abrupt, or otherwise unnatural breath-like segments (e.g., E\_0002422651, E\_0006333696) are highlighted with high spoof attributions, reflecting flawed breath synthesis in deepfake synthesizers. Unfortunately, the models can be tricked by deepfakes that convincingly mimic respiratory noise.

\textbf{Vulnerability to Audio Compression}: A critical vulnerability shared across all three models is severe degradation under heavy audio compression, which was highly prevalent in the \textit{High-confidence errors} subset. Heavily compressed bona fide recordings (e.g., AMR 8kHz or Opus codecs) were consistently assigned high spoof scores. For example, in compressed bona fide files, annotators noted that the speech was ``missing depth" and that ``energy is flat even in strong phonemes" (e.g., E\_0009198285), and the models flagged codec artifacts across the entire recording. This reveals a shared weakness: the models incorrectly treat the compression artifacts as those of deepfake synthesis, as shown in \autoref{tab:validation}.

\begin{table}[tbp]
\centering
\footnotesize
\caption{Performance on ASVspoof~5 evaluation set. LR fusion denotes a logistic regression score fusion of the three detector scores trained on the development set.}
\label{tab:fusion_results}
\begin{tabular}{l|ccc|c}
\toprule
\textbf{Metric} & \textbf{AASIST} & \textbf{CA-MHFA} & \textbf{SLS} & \textbf{LR Fusion} \\
\midrule
\textbf{EER (\%)}  & 4.06\% & 5.26\% & 3.98\% & 3.77\% \\
\textbf{minDCF}    & 0.1015 & 0.1330 & 0.1040 & 0.0970 \\
\bottomrule
\end{tabular}
\vspace{-1em}
\end{table}

\subsection{Validation}

By ablating attributes that appear to drive the detector, we aim to experimentally validate the observed phenomena. We apply each operation to the full eval set and report the performance shifts in \autoref{tab:validation}:

\textbf{Silence masking}: Using a Wav2Vec2 forced-aligner~\cite{Baevski_wav2vec2}, we detect non-speech frames and mix their SSL embeddings with a bona fide silence centroid as $h_{m} = 0.1 h_{sil} + 0.9\boldsymbol{\mu}_{sil}$ where $h_m, h_{sil}$ are the masked and detected non-speech frame, respectively, and $\boldsymbol{\mu}_{sil}$ is the bona fide silence centroid\footnote{The centroid is computed similarly as $\boldsymbol{\mu}$ in \autoref{subsec:baseline}, but with silence from bona fide training recordings only.}. This ensures that we do not introduce auxiliary artifacts through hard cuts or out-of-domain replacements (see \autoref{subsec:baseline}).

\textbf{High-energy phoneme masking}: We use the same forced-aligner to mark high-energy phoneme frames (above 75\% energy range of the recording) and replace each marked SSL frame by the mean of its nearest unmasked neighbors, removing localized content while preserving in-domain continuity.

\textbf{Spectral masking}: To test frequency-specific cues, we apply an STFT with a Hann window and reduce the energy of the frequency band between 1000Hz and 1600Hz to 10\% of the original energy, followed by inverse STFT reconstruction.


\textbf{Compressor effect}: To test sensitivity to energy dynamics, we attenuate amplitudes exceeding -$20$\,dB and subsequently add $10$\,dB gain to the entire recording. This effectively reduces the dynamic range, suppressing high-energy peaks and boosting quieter segments and background noise.

Results in \autoref{tab:validation} experimentally complement the manual annotations\footnote{$\text{FAR}_{b}, \text{FRR}_{b}$ are computed under a fixed baseline EER threshold.}. \textit{Silence} masking catastrophically impacts AASIST ($\text{FAR}_{b}$ jumps to 99.99\%, strongly shifting scores toward bona fide) and heavily degrades SLS ($\text{FAR}_{b}$ 47\%), while leaving CA-MHFA relatively unaffected. This directly validates the findings that AASIST evaluates the environmental noise floor and unvoiced regions. CA-MHFA immunity to this masking confirms it primarily hunts for artifacts within articulated speech rather than evaluating non-speech pauses.


\textit{High-energy phoneme masking} massively derails AASIST (EER jumping to 17.81\%, with $\text{FAR}_{b}$ at 22.41\%) while CA-MHFA and SLS remain largely robust. Replacing high-energy frames with smoothed neighborhoods introduces a messy effect into the signal. Because AASIST operates as a global environmental anomaly detector, it flags this artificial, non-reverberant edit as a massive spoofing artifact. Conversely, the smoothing operation replaces spiky glitches that SLS disregards (only a~small change in EER), but it also likely shifts CA-MHFA's focus to other unmasked consonants and word boundaries.

\textit{Spectral masking} slightly degrades all detectors, consistent with our observations: SLS focuses on the missing mid-energy frequency band, CA-MHFA relies on missing formants or phoneme integrity, and AASIST is sensitive to the environment anomaly created by the masking.

Finally, the \textit{Compressor} ablation significantly degrades all three models. The directional error shift is heavily concentrated in false rejections ($\text{FRR}_b$ spikes to 8.38-11.93\% while $\text{FAR}_b$ remains relatively stable), i.e., more spoof-like scores. By boosting the background noise floor and attenuating peak speech dynamics, the compressor computationally mimics the lossy codec artifacts observed in the manual a\-naly\-sis. This confirms our hypothesis that the models universally connect flattened energy and diffuse codec noise to synthesis artifacts, thereby more likely flagging bona fide human speech as a deepfake.

\begin{table}[tbp]
    \footnotesize
    \centering
    \caption{Ablation results (percentages \%) on the ASVspoof~5 evaluation set. FAR$_b$ (deepfake accepted as bona fide) and FRR$_b$ (bona fide flagged as deepfake) are evaluated at the fixed decision threshold derived from the Baseline EER to illustrate the directional shift in scores.}
    \label{tab:validation}
    \setlength{\tabcolsep}{2pt}
    \begin{adjustbox}{max width=\linewidth}
    \begin{tabular}{@{}l rrr|rrr|rrr}
        \toprule
        \multirow{2}{*}{\textbf{Modif.}} 
          & \multicolumn{3}{c}{\textbf{AASIST}}
          & \multicolumn{3}{c}{\textbf{CA-MHFA}}
          & \multicolumn{3}{c}{\textbf{SLS}} \\
        \cmidrule(lr){2-4}\cmidrule(lr){5-7}\cmidrule(lr){8-10}
          & \textbf{EER} & \textbf{$\text{FAR}_{b}$} & \textbf{$\text{FRR}_{b}$} 
          & \textbf{EER} & \textbf{$\text{FAR}_{b}$} & \textbf{$\text{FRR}_{b}$}
          & \textbf{EER} & \textbf{$\text{FAR}_{b}$} & \textbf{$\text{FRR}_{b}$} \\
        \midrule
        \rowcolor{black!8} 
        \textbf{Baseline}   & 4.06 & 4.06 & 4.06 & 5.26 & 5.26 & 5.26 & 3.98 & 3.98 & 3.98 \\
        \textbf{Silence}    & 5.08 & 99.99 & 0.01 & 5.32 & 5.50 & 5.19 & 4.54 & 47.01 & 0.90 \\
        \textbf{Phoneme}    & 17.81 & 22.41 & 8.73 & 5.17 & 5.12 & 5.20 & 4.03 & 5.64 & 3.20 \\
        \textbf{Spectral}   & 4.53 & 3.97 & 4.48 & 5.55 & 4.83 & 6.05 & 4.53 & 4.30 & 4.52 \\
        \textbf{Compr.}     & 6.78 & 4.61 & 8.38 & 7.67 & 3.86 & 11.93 & 7.32 & 4.18 & 9.82 \\
        \bottomrule
    \end{tabular}
    \end{adjustbox}
    \vspace{-1em}
\end{table}

\section{Conclusion}

We introduce an audio-native explainability pipeline that moves beyond speculative reasoning and empirically demonstrates which specific cues the detectors focus on. Through rigorous manual annotations and causal experimental validation, we identify the primary cues of the examined detectors and assign them semantic meaning: 
\begin{itemize}
    \item \textbf{AASIST} acts as an environmental anomaly detector, relying heavily on non-speech regions.
    \item \textbf{CA-MHFA} functions as a highly localized detector targeting specific articulation and phoneme artifacts.
    \item \textbf{SLS} monitors overall spectral integrity, temporal continuity of speech, and word boundaries.
\end{itemize}

Despite the differences, the examined detectors share a vulnerability to heavy audio compression, incorrectly treating the flattened energy and diffuse noise of lossy codecs as synthetic generation artifacts. Ultimately, because their learned representations and primary cues do not significantly overlap, the detectors operate as complementary observers. This lack of consensus suggests that ensemble methods could benefit from combining these architectures, although shared failure modes such as heavy compression constrain the achievable gains without targeted mitigation. Our findings enable informed detector design choices, can support forensic audio analyses, and help anticipate detector behavior in broad deployment scenarios.

\section{Acknowledgments}

This work was supported by the Brno University of Technology internal project FIT-S-26-9011. Computational resources were provided by the e-INFRA CZ project (ID:90254), supported by the Ministry of Education, Youth and Sports of the Czech Republic.

\section{Generative AI Use Disclosure}

During the preparation of this work, the authors used Generative AI Models (specifically Google Gemini, ChatGPT, and Grammarly) for language editing and text refinement. The authors reviewed and edited the output as needed and take full responsibility for the publication's content.

\bibliographystyle{IEEEtran}
\bibliography{mybib}

\end{document}